\documentclass[prl,aps,twocolumn,showpacs,superscriptaddress]{revtex4}
\usepackage{amsbsy}
\usepackage{bm}
\usepackage{graphicx}

\newcommand{\nc}{\newcommand}
\nc{\noi}{\noindent}     
\nc{\eq}[1]{\mbox{Eq.~(\ref{#1})}}
\nc{\ba}{\begin{array}}
\nc{\ea}{\end{array}}
\nc{\bea}{\begin{eqnarray}}
\nc{\eea}{\end{eqnarray}}
\nc{\fig}[1]{\mbox{Fig.~\ref{#1}}}

\begin{document}

\title{Reply to Comment on  `Spin Decoherence in Superconducting Atom Chips'}

\author{Bo-Sture K. Skagerstam}\email{boskag@phys.ntnu.no}
\affiliation{Complex Systems and Soft Materials Research Group, Department of Physics, 
             The Norwegian University of Science and Technology, N-7491 Trondheim, Norway}
\author{Ulrich Hohenester}
\affiliation{Institut f\"ur Physik, Karl-Franzens-Universit\"at Graz, Universit\"atsplatz 5, A-8010 Graz, Austria}
\author{Asier Eiguren}
\affiliation{Institut f\"ur Physik, Karl-Franzens-Universit\"at Graz, Universit\"atsplatz 5, A-8010 Graz, Austria}
\author{Per Kristian Rekdal}
\affiliation{Institut f\"ur Physik, Karl-Franzens-Universit\"at Graz, Universit\"atsplatz 5, A-8010 Graz, Austria}


\maketitle

     In a recent paper \cite{skagerstam_06} we investigate spin decoherence in superconducting atom chips,
     and predict a lifetime enhancement by more than five orders of magnitude in comparison to
     normal-conducting atom chips. Scheel, Hinds, and Knight (SHK) \cite{scheel_06} cast doubt on these
     results as they are seemingly an artifact of the two-fluid model used for the description of the superconductor,
     and estimate a lifetime enhancement by a factor of ten instead. In this reply we show that this criticism
     is unwarranted since neither our central result relies on the two-fluid model, nor the predictions of our
     model strongly disagree with experimental data.

    In Ref.~\cite{skagerstam_06} we employ a dielectric description of the superconductor based on a parameterization
    of the complex optical conductivity $\sigma(T) \equiv \sigma_1(T) + i \sigma_2(T)$, viz.

\begin{equation}   \label{sigma_eq}
  \sigma(T) = 2/\omega\mu_0\delta^2(T)+i/\omega\mu_0\lambda_L^2(T) \, ,
\end{equation}

    \noi
    with $\delta(T)$ and $\lambda_L(T)$ the temperature dependent skin and London penetration depth, respectively.
    The spin lifetime for $\lambda_L(T) \ll \delta(T)$ is then obtained by matching the electromagnetic fields at
    the vacuum-superconductor interface, to arrive at our central result for the spin lifetime
    $\tau \propto \sigma^{3/2}_2(T)/\sigma_1(T) \propto \delta^2(T)/\lambda_L^3(T)$. As our analysis is only based
    on Maxwell's theory with appropriate boundary conditions, it is valid for $\delta(T)$ and $\lambda_L(T)$
    values obtained from either a microscopic model description or from experimental data on $\sigma(T)$.
    The specific choice of parameterization in \eq{sigma_eq} is motivated by the
    two-fluid model, even though this model is not needed to justify it. 
    %

    In order to obtain an estimate of the lifetime for niobium, we make use of the experimental value $\sigma_1(T_c) = \sigma_n$
    \cite{casalbuoni_06} and consider for $T \leq T_c$ the Gorter-Casimir temperature dependence
    $\sigma_1(T) = (T/T_c)^4 \sigma_n$ and $\sigma_2(T) = ( 1 - (T/T_c)^4 ) \, \sigma_2(0)$, 
    where $\sigma_2(0) = 1/\omega \mu_0 \lambda_L^2(0)$.
    A decrease in temperature from $T_c$ to $T_c/2$ as considered in
    Ref.~\cite{scheel_06}, then results in a reduction of $\sigma_1(T_c/2)$ by approximately one order of magnitude.
    SHK correctly note that the modification of the quasi-particle dispersion in the superconducting state might
    give rise to a coherence Hebel-Schlichter peak of $\sigma_1(T)$ below $T_c$ \cite{lifetime}.
    To estimate the importance of this peak, we have computed $\sigma_1(T)$ using the Mattis-Bardeen formula.
    At the atomic transition frequency of 560 kHz we obtain a peak height of approximately $5 \, \sigma_n$,
    not hundred $\sigma_n$ \cite{scheel_05} as used by SHK.
    From the literature it is well-known that this coherence peak becomes substantially reduced if lifetime effects
    of the quasi-particles are considered \cite{lifetime,lifetime_2} and even disappears in the clean superconductor limit \cite{marsiglio:91}.
    As a fair and conservative estimate we correspondingly assign an uncertainty of one order of magnitude to our spin lifetimes.
    On the other hand, the major contribution of the $\tau$ enhancement in the superconducting state is due to the additional
    $\lambda_L^3(T)$ contribution accounting for the efficient magnetic field screening in superconductors.
    This factor is not considered by SHK and appears to be the main reason for the discrepancy between our results and those
    of Ref.~\cite{scheel_06}. 
    The London length $\lambda_L(0) = 35$ nm  as used in
    \cite{skagerstam_06} corresponds to $\omega \sigma_2(T_c/2) \approx 6.1 \times 10^{20} \, ( \Omega \, $m$ \, $s$ )^{-1}$
    and is in agreement with the experimental data of Ref.~\cite{casalbuoni_06}.
    %

    In conclusion, modifications of $\sigma(T)$ introduced by the details of the quasi-particle dispersion in the
    superconducting state (BCS or Eliashberg theory) are expected to modify the estimated lifetime values by at
    most one order of magnitude, but will by no means change the essentials of our findings, which only rely on
    generic superconductor properties. Hence, our prediction for a lifetime enhancement by more than five orders
    of magnitude prevails. Whether such high lifetimes can be obtained in superconducting atom chips
    will have to be determined experimentally.



\begin{thebibliography}{xx}





   \bibitem{skagerstam_06}
   B.-S.K. Skagerstam, U. Hohenester, A. Eiguren, and P.K. Rekdal,
   Phys. Rev. Lett. {\bf 97}, 070401 (2006).     


   \bibitem{scheel_06}
   S. Scheel, E. Hinds, and P.L. Knight, arXiv:quant-ph/0610095 (2006).






     \bibitem{casalbuoni_06}
     S. Casalbuoni, E.A. Knabbe, J. K\"otzler, L. Lilje, L. von Sawilski, P. Schm\"user, and B. Steffen, 
     Nucl. Instrum. Methods Phys. Res. A {\bf 538}, 45 (2005).   



     \bibitem{lifetime}
     L. C. Hebel and C. P. Slichter, Phys. Rev. {\bf 113}, 1504 (1959);
     O. Klein {\em et al.},\/ Phys. Rev. B {\bf 50}, 6307 (1994).


     \bibitem{scheel_05}
     S. Scheel, P.K. Rekdal, P.L. Knight, and E.A. Hinds,
     Phys. Rev. A {\bf 70}, 042901 (2005).



     \bibitem{lifetime_2}
     F. Marsiglio {\em et al.},\/ Phys. Rev. B {\bf 50}, 7203 (1994).


     \bibitem{marsiglio:91}
     F. Marsiglio, Phys. Rev. B {\bf 44}, 5373 (1991).

  





\end{thebibliography}
\end{document}